\documentstyle[epsfig]{aipproc} 

\begin{document}

\title{Charge Conjugation Violation in Supernovae and The Neutron Shortage for R-Process Nucelosynthesis}

\author{C. J. Horowitz and Gang Li} 
\address{Nuclear Theory Center and Dept. of Physics,
Indiana University, Bloomington, IN 47405}

\maketitle 
\begin{abstract}
Core collapse supernovae are dominated by energy transport from
neutrinos.  Therefore, some supernova properties could depend
on symetries and features of the standard model weak interactions.  
The cross section
for neutrino capture is larger than that for antineutrino capture by one term 
of order the neutrino energy over the nucleon mass.  This reduces the ratio of 
neutrons to protons in the $\nu$-driven wind above a protoneutron star 
by approximately 20 \% and may significantly hinder r-process nucleosynthesis.
\end{abstract}

Core collapse supernovae are perhaps the only present day large systems
dominated by the weak interaction.   They are so dense that photons and
charged particles diffuse very slowly.   Therefore energy transport is by
neutrinos (and convection).

We beleive it may be useful to try and relate some supernova properties 
to the symmetries and features of the standard model weak interaction. 
Parity violation in a strong magnetic field could lead to an asymmetry of
the explosion\cite{1}.  Indeed, supernovae explode with a dipole asymmetry of
order one percent in order to produce the very high `recoil' velocities
observed for  neutron stars\cite{2}.   However, calculating the expected
asymmetry from P violation has proved complicated.  Although explicit
calculations have yielded somewhat small asymmetries\cite{3,4,5} it is still
possible that more efficient mechanisms will be found.

In this paper we calculate effects of charge conjugation, C, violation in the Standard Model on the difference between neutrino and antineutrino interactions.  In Quantum Electrodynamics C symmetry insures the cross section for $e^-p$ is equal to that for $e^+p$ scattering 
(to lowest order in $\alpha$).  In contrast, the standard model has large parity, P, and C violations (since the product CP is approxamitely conserved).  Therefore the $\bar\nu$-nucleon cross sections are systematically smaller than $\nu$-nucleon cross sections.

However at the low $\nu$ energies in supernovae, time reversal symmetry
limits the difference between $\nu$ and $\bar\nu$ cross sections.  
Time reversal can relate $\nu-N$ elastic scattering and $\bar\nu-N$ where 
the nucleon scatters from final momentum $p_f$ to initial momentum $p_i$.  
If the nucleon does not recoil then the $\nu$ and $\bar\nu$ cross sections 
are equal.   Thus the difference between $\nu$ and $\bar\nu$ cross sections 
are expected to be of recoil order $E/M$ where $E$ is the neutrino energy 
and $M$ the nucleon mass.  This ratio is relatively small in
supernovae.  However the coefficient multiplying $E/M$ involves the large
weak magnetic moment of the nucleon (see below).

The standard model has larger $\nu$ cross sections than those for $\bar\nu$.
For neutral currents, this leads to a longer mean free path for $\bar\nu_x$ 
compared to $\nu_x$ (with x=$\mu$ or $\tau$).  Thus even 
though $\nu_x$ and $\bar\nu_x$ are produced in pairs, the antineutrinos 
escape faster leaving the star neutrino rich.  The muon and tau number 
for the protoneutron star in a supernova could be of order $10^{54}$\cite{6}.  
Supernovae may be the only known systems with large $\mu$ and or $\tau$ number.
For charged currents, the interaction difference can change the 
equilibrium ratio of neutrons to protons and may have important 
implications for nucleosynthesis.  We discuss this below.  To our knowledge, 
all previous work on nucleosynthesis in supernovae assumed equal $\nu$ 
and $\bar\nu$ interactions (aside from the n-p mass difference).

The neutrino driven wind outside of a protoneutron star is an attractive
site for r-process nucleosynthesis\cite{7}.  Here nuclei rapidly capture neutrons
from a low density medium to produce heavy elements\cite{8}.    This requires,
as a bare minimum, that the initial material have more neutrons than protons.
The ratio of neutrons to protons n/p in the wind depends on the rates for
the two reactions:
$$\nu_e + n \rightarrow p + e^-,\eqno(1a)$$
$$\bar\nu_e+p \rightarrow n + e^+.\eqno(1b)$$
The standard model cross sections for Eqs. (1a,1b) to order $E/M$ are,
$$\sigma={G^2{\rm cos}^2\theta_c\over \pi}(1+3g_a^2) E_e^2
[1-\gamma {E\over M} \pm \delta{E\over M}],\eqno(2)$$
with $G$ the Fermi constant (and $\theta_c$ the Cabbibo angle),
$E_e=E\pm \Delta$ the energy of the charged lepton and $\Delta=1.293$ MeV is
the neutron-proton mass difference.  The plus sign is for Eq. (1a) and the
minus sign for Eq. (1b).  We use $g_a\approx 1.26$. Equation (2) neglects 
small corrections involving the electron mass and coulomb effects while the 
finite nucleon size only enters at order $(E/M)^2$.

We refer to the $\gamma$ term as a recoil correction.  It is the same for
$\nu$ and $\bar\nu$, $\gamma=(2+10g_a^2)/(1+3g_a^2)\approx 3.10$.
Finally, the $\delta$ term, $\delta=4g_a(1+2F_2)/(1+3g_a^2)\approx 4.12$,
involves the interference of vector (1+2$F_2$) and 
axial ($g_a$) currents.  This violates P, which by CP invariance also violates
C.   This increases the $\nu$ and decreases the $\bar\nu$ cross section,
Note, $F_2$ is the isovector anomalous moment of the nucleon. (This is the
weak magnetism contribution.)

The equilibrium electron fraction per baryon $Y_e$ (which is equal to the
proton fraction assuming charge neutrality) is simply related to the rate
$\bar\lambda$ for Eq. (1b) divided by the rate $\lambda$ for Eq. (1a).
$$Y_e=(1+{\bar\lambda\over \lambda})^{-1}\eqno(3)$$
The ratio neutrons to protons is, ${n\over p} = {1\over Y_e} -1$.

In ref.~\cite{prl} we calculate the reaction rates by averaging Eq. (2) over neutrino spectra to get,
$$Y_e=\Bigl(1+{L_{\bar\nu_e}\bar\epsilon\over L_{\nu_e}\epsilon}
QC\Bigr)^{-1}.\eqno(4)$$
Here $\epsilon$ ($\bar\epsilon$) is the $\nu_e$ ($\bar\nu_e$) mean energy, $L_{\nu_e}$ ($L_{\bar\nu_e}$) is the $\nu_e$ ($\bar\nu_e$) luminosity,
$Q$ is the correction from the reaction Q value,
$$Q={1-2{\Delta\over \bar\epsilon}+a_0{\Delta^2\over \bar\epsilon^2}\over
1+2{\Delta\over \epsilon}+a_0{\Delta^2\over\epsilon^2}},\eqno(5)$$
and with C violating one has a factor $C$,
$$C={1-(\delta+\gamma)a_2{\bar\epsilon\over M}\over
1+(\delta-\gamma)a_2{\epsilon\over M}}.\eqno(6)$$
Simply evaluating Eq. (6) for typical parameters yields $C\approx 0.8$.  
Thus, {\it the difference between $\nu$ and $\bar\nu$ interactions reduces the 
equilibrium n/p ratio by approximately 20 \%.}  This is an important result and 
will be discussed below.

Evaluating Eq. (4) for the neutrino fluxes of a Supernova simulation by Wilson~\cite{11} shows that with the $C$ term the neutrino driven wind starts out proton rich and ends up with about equal numbers of neutrons and protons.  When C violation is included the wind is never significantlt neutron rich.  It is very unlikely that successful r-process nucleosynthesis can
take place in the wind of this or similar models.

With the approximately 20 \% reduction in n/p from the difference 
between $\nu$ and $\bar\nu$ interactions, there appears to be 
very serious problems with r-process nucleosynthesis in the
wind of present supernova models.  In addition to the initial lack of
neutrons,  one has to overcome the effects of neutrino interactions during
the assembly of $\alpha$ particles and during the r-process itself\cite{14}.
These further limit the available neutrons per seed nucleus.  Thus, it is
unlikely that present wind models will produce a successful r-process.  Of
course, the wind in supernovae may not be the r-process site, although this
may be unappealing (see for example\cite{8,15}).  If the wind is not the site,
one must look for alternative environments.

However, the effects of neutrino interactions may be very general.  The only 
requirement is that energy transport from neutrinos plays some role in helping 
material out of a deep gravitational well.   Given this, it is quite likely 
that the n/p ratio will be determined by the relative rates of Eqs. (1a,1b).  
Therefore differences in $\nu$ and $\bar\nu$ interactions may be important for 
just about any nucleosynthesis site that involves neutrinos.

If the $\nu$-driven wind is the r-process site, it is very likely,
present models of the neutrino radiation in supernovae are incomplete.  The
high values of $Y_e$ make it almost impossible to have a successful r-process
by only changing matter properties, such as the entropy.  The neutrino fluxes
will (almost assuredly) need to be changed.

Changes in the astrophysics used in the simulations or new neutrino physics
such as neutrino oscillations\cite{17} could change $\bar\epsilon$, $\epsilon$
and or the luminosities and lead to a more neutron rich wind.  The
oscillations of more energetic $\bar\nu_x$ with $\bar\nu_e$ could increase
$\bar\epsilon$.  However, we have some information on the $\bar\nu_e$
spectrum from SN1987a\cite{18}.  Thus one can not increase $\bar\epsilon$ 
without limit.  Indeed if anything, the Kamiokande data suggest a lower
$\bar\epsilon$.  Any model which tries to solve r-process nucleosynthesis
problems by increasing $\bar\epsilon$ should first check consistency with
SN1987a observations\cite{19}.  Alternative modifications could include
oscillations of $\nu_e$ to a sterile neutrino or a {\it lowering} of
$\epsilon$.  (However, we know of no model which lowers $\epsilon$.)
Whatever the modification of the neutrino fluxes, one will still need to
include the differences between $\nu$ and $\bar\nu$ interactions
in order to accurately calculate n/p.

In conclusion, supernovae are one of the few large systems dominated by
energy transport from weakly interacting neutrinos.  Therefore, some 
supernova properties may depend on symmetries and features
of the standard model weak interactions.  The cross secton for neutrino
capture is larger than that for antineutrino capture by a term of
order the neutrino energy over the nucleon mass.  This difference between 
neutrino and antineutrino interactions reduces the ratio of neutrons 
to protons in the $\nu$-driven wind above a protoneutron star by 
approximately 20 \% and may significantly hinder r-process 
nucleosynthesis.

This work was supported in part by DOE grant: DE-FG02-87ER40365.

\end{document}